\documentclass[11pt]{JHEP3}

\usepackage{epsfig}
\epsfclipon
\usepackage{multicol}
\usepackage{epsfig,bm}
\usepackage{amssymb,amsmath}

\newcommand{\roughly}[1]{\mathrel{\raise.3ex\hbox{$#1$\kern-0.85em
\lower1ex\hbox{$\sim$}}}}

\newcommand{\lsim}{\roughly<}
\newcommand{\gsim}{\roughly>}

\def\be{\begin{equation}}
\def\beq\begin{equation}
\def\ee{\end{equation}}
\def\bea{\begin{eqnarray}}
\def\eea{\end{eqnarray}}

\def\beq{\begin{equation}}
\def\eeq{\end{equation}}
\def\beqa{\begin{eqnarray}}
\def\eeqa{\end{eqnarray}}

\newcommand{\bmat}{\left(\begin{array}}
\newcommand{\emat}{\end{array}\right)}

\def\nisubsubsection#1{\medskip\noindent {\bf #1} \smallskip\noindent}

\def\yzero{\smash{\hbox{$y\kern-4pt\raise1pt\hbox{${}^\circ$}$}}}

\def\-{\hphantom{-}}

\def\s2{\frac{1}{2}}

\def\IF{\relax{\rm I\kern-.18em F}}
\def\II{\relax{\rm I\kern-.18em I}}
\def\IP{\relax{\rm I\kern-.18em P}}
\def\IC{\relax{\rm I\kern-.48em C}}
\def\IR{\relax{\rm I\kern-.18em R}}
\def\IK{\relax{\rm I\kern-.20em K}}
\def\IM{\relax{\rm I\kern-.25em M}}

\def\Dsl{\,\raise.15ex\hbox{/}\mkern-13.5mu D} %this one can be subscripted
\def \one{\relax{\rm 1\kern-.26em I}}

\title{MSLED~: A Minimal Supersymmetric Large Extra Dimensions Scenario}

\author{C.P.~Burgess,$^1$ J.~Matias\phantom{,}$^2$ and
F.~Quevedo\phantom{,}$^3$
\\

$^1$ Physics Department, McGill University,  3600 University
Street, Montr\'eal, Qu\'ebec, Canada, H3A 2T8. \\

$^2$ IFAE, Universitat Aut\`onoma de Barcelona, 08193 Bellaterra, Barcelona, Spain.\\

$^3$ Centre for Mathematical Sciences, DAMTP,
               University of Cambridge,\\
               Cambridge CB3 0WA UK.}

\date{}
\abstract{We propose a framework for the low-energy realization of
supersymmetry which is very predictive, but differs radically in
its phenomenological implications from the supersymmetric Standard
Model (minimal or otherwise). The proposal consists of a
supersymmetric version of the Large-Extra-Dimensions scenario,
with the Standard Model living on a 3-brane, coupled to a bulk
sector consisting of six-dimensional supergravity. This picture is
motivated by a promising recent attempt ({\tt hep-th/0304256}) to
naturally understand the observed dark energy density, and this
connection with dark energy prevents making the extra dimensions
smaller than of order 5 $\mu$m. The resulting inability to change
this size makes the model very predictive, and easily falsifiable
within the near future. Being supersymmetric, it may plausibly be
embedded into a more fundamental theory such as string theory, in
which case an additional 4 compact dimensions may also be present
having inverse  radii at  the TeV scale or higher. The model is
close to, but consistent with, current experimental constraints.
We outline possible phenomenological implications for particle
physics (both at accelerators and elsewhere), for precision tests
of gravity, for astrophysics and for cosmology.}

\preprint{McGill-04/08, UAB-FT-564, DAMTP-2004-38}

\keywords{Strings, Branes, Cosmology}

\begin{document}

%\newpage

%===================================================================================

\section{Introduction}

If supersymmetry were unbroken it would solve both the hierarchy
and the cosmological constant problems, in the sense that the
bose-fermi cancellations implied by unbroken supersymmetry could
keep ultra-violet effects from generating dangerous
contributions to the Higgs boson mass and to the vacuum energy.
Unfortunately, the failure to find super-partners for the observed
elementary particles appears to imply that the effective
supersymmetry breaking scale is at least of order 1 TeV.
Consequently, in practice low-energy supersymmetry is used only to
solve the hierarchy problem, sacrificing its potential for solving
the cosmological constant problem. This is true for essentially
all of the proposed supersymmetric extensions of the Standard
Model --- including the minimal (MSSM) version.

This line of thinking is reasonably compelling within the standard
4-dimensional setting, but more possibilities may exist within the
framework of the brane-world scenario with large extra dimensions
(LED). Indeed, recently there has been progress towards
understanding how the small size of the observed Dark Energy
\cite{ccnonzero} --- $\lambda_{\rm obs} \approx (0.003$ eV$)^4$
--- can naturally emerge from theories having supersymmetric large
extra dimensions (SLED) \cite{Towards,Warped,Update}. Because
these theories invoke new physics at sub-eV energies, they are
likely to have many striking implications for cosmology, particle
physics (both at colliders and at lower energies) and precision
tests of general relativity.

In this article we wish to begin the discussion of the main
observational implications of these theories, specializing where
necessary to a minimal version of the SLED proposal (which we call
MSLED). Besides being well-motivated from the
cosmological-constant point of view, this minimal version has the
great virtue of being very predictive. In particular, its use to
explain the size of the observed Dark Energy removes the freedom
to `move the goalposts' by significantly changing the number and
size of the extra dimensions, and so is much more easily falsified
than is often true for extra-dimensional proposals.

\subsection{Why SLED?}

The SLED proposal \cite{Towards,Update} --- like the LED proposal
before it \cite{ADD} --- posits that at present there are two
extra dimensions whose circumference, $r$, is of order $r \sim 10
\, \mu\hbox{m} \sim (10^{-2}$ eV$)^{-1}$. This is only possible if
all of the known particles and interactions besides gravity are
trapped on a 3-brane which sits at a point within these extra
dimensions, and if the scale of gravitational physics in the extra
dimensions is of order $M_g \sim 10$ TeV.

SLED also supposes these large extra dimensions arise within a
supersymmetric field theory, such as might be expected to arise in
the low-energy limit of string theory. This can only work if
supersymmetry is badly broken on our 3-brane, since we know that
there are no super-partners for the observed particles having
masses which are much smaller than $M_g$. Given this scale for
supersymmetry breaking on the brane there is also a trickle-down
of supersymmetry breaking to the `bulk' between the branes, whose
size is set by the bulk's Kaluza-Klein scale and so which can be
as low as $m_{sb} \sim 10^{-2}$ eV.\footnote{The supersymmetry
breaking scale would be larger if the extra dimensions should be
warped.}

Within this framework gravitational physics is effectively
6-dimensional for any energies above the scale, $1/r \sim 10^{-2}$
eV, and so the cosmological constant problem must be posed within
this 6-dimensional context. In 6 dimensions the vacuum energy due
to all of the known particles is localized on our brane, and so is
really a localized energy source in the extra dimensions rather
than directly contributing a 4-dimensional cosmological constant.
Einstein's equations then dictate how the extra dimensions curve
in response to this local energy distribution, with the result
that the resulting curvature precisely cancels (at the classical
level) the vacuum energy on the branes \cite{CLP}. If the bulk is
supersymmetric then a similar classical cancellation also occurs
amongst the other bulk contributions to the effective 4D
cosmological constant \cite{Towards,Warped,GGP}.

Quantum effects in the bulk ruin this perfect cancellation of the
4D vacuum energy, but by an amount which is controlled by the
supersymmetry-breaking scale within the bulk ({\it i.e.} by
$m_{sb} \sim 10^{-2}$ eV). Although for some theories these
corrections can be of order $\lambda \sim m_{sb}^2 M_g^2$, for
others this leading contribution also cancels, leaving a residual
result $\lambda \sim m_{sb}^4$, which is the right order of
magnitude to describe the observed density of Dark Energy
\cite{Towards,Update,Doug}.

Some open issues remain about whether this proposal provides a
completely satisfactory mechanism for naturally obtaining a
sufficiently small cosmological constant. Most notable among these
is the issue of whether the internal space should prefer to
dynamically warp and so to raise the scale $m_{sb}$ without
changing $M_p$ \cite{Update}. This is almost certain to happen
given the comparatively large energies available in the earlier
universe, making it an interesting dynamical question whether this
warping should persist into the present epoch. Even should it do
so, however, the SLED proposal provides a mechanism whereby the
question of the smallness of $\lambda$ becomes a dynamical issue,
and represents a step forward relative to previous proposals.
Furthermore, it provides a completely new way in which
supersymmetry might be realized at low energies. For these reasons
we believe it provides a very well-motivated framework whose
phenomenological consequences are worth exploring in their own
right.

\subsection{What is MSLED?}

Because the SLED proposal modifies physics at sub-eV scales, it
has many other observational implications besides its explanation
for the small size of the cosmological constant. These
implications are most restrictive if they are cast in terms of a
particularly simple variant of the SLED proposal which, following
\cite{Update}, we call `minimal' SLED or MSLED.

The action for any SLED variant has the form
\beq
    S_{\rm tot} = S_{SG} + S_{3} + S_{3'} + S_{\rm int} \,,
\eeq
where $S_{SG}$ describes the action of 6-dimensional supergravity,
$S_3$ contains the physics of the 3-brane on which we find
ourselves situated, $S_{3'}$ describes the physics of any other
parallel branes and $S_{\rm int}$ describes the interactions
between our brane and the 6D supergravity fields in the bulk.
MSLED makes the following particularly simple choices for $S_{SG}$
and $S_3$.

\medskip\noindent {\it Our Brane:}
We first choose the degrees of freedom on our brane to be given
simply by the usual Standard Model particle content, including its
elementary Higgs field. Since the Standard Model can be defined as
the most general renormalizable theory of this particle content,
we are guaranteed that the most general renormalizable
interactions which can be included in $S_3$ is simply the Standard
Model itself. Furthermore, since we are dealing with a low-energy
effective theory (as we must for any field theory of gravity) we
also expect higher-dimensional, non-renormalizable interactions to
arise in $S_3$ suppressed by powers of $M_g$. Since the
gravitational scale of the 6D physics is $M_g \sim 10$ TeV in the
SLED picture, with a few exceptions these higher-dimensional
effective interactions are generically small enough to have
escaped detection until now.\footnote{The few exceptions are those
interactions which violate symmetries like lepton number or baryon
number, which we must envision as being forbidden by conservation
laws. More about this later.} Although the assumption of this
minimal particle content is not required in order to have a small
cosmological constant, it is by far the most conservative
possibility whose restrictive predictions should first be
explored.\footnote{There are also two kinds of non-standard brane
fields which may also be required on our brane in addition to the
Standard Model. These are the Goldstone modes associated with the
position of our brane in the extra dimensions, and with the
breaking of supersymmetry on our brane. We do not follow these
modes explicitly here because they get eaten by bulk modes, and so
are natural to include with the bulk-brane interactions, $S_{\rm
int}$.}

\medskip\noindent {\it The Bulk:}
The physics of the bulk, $S_{SG}$, is described by 6D
supergravity. Since 6D supergravity comes in several varieties,
this last choice contains several sub-options. As we shall see,
observational constraints restrict the number of degrees of
freedom which can live in the bulk, and so disfavor chiral
supergravities like Nishino-Sezgin supergravity \cite{NS}. They
instead point towards the non-chiral, ungauged versions of 6D
supergravity, for which the field content contains only a minimal
particle content, such as consisting of a scalar dilaton ($\phi$),
two symplectic majorana-weyl spin-1/2 dilatini ($\chi^r, r=1,2$),
2-form gauge potential ($B_{MN}$), two symplectic majorana-weyl
gravitini ($\psi_M^r$) and metric ($g_{MN}$).\footnote{Strictly
speaking, this multiplet is reducible, since a smaller gravity
multiplet can be built using only the metric, gravitino and the
self-dual piece of $B_{MN}$.} Chiral and non-chiral models differ
in whether or not the two bulk dilatini, $\chi$, (or the two bulk
gravitini, $\psi_M$) share the same 6D chirality: $\Gamma_7 \chi^r
= \pm \chi^r$.

\medskip

It must be emphasized that these choices differ considerably in
their implications for upcoming experiments from the presently
popular paradigm of the supersymmetric Standard Model, in its
minimal (MSSM) and more complicated variants. In particular, the
assumption that the physics of our brane is described by the
Standard Model, brings back all of the virtues of this model which
had to be sacrificed when the MSSM was adopted. This allows us to
retain a natural understanding as to why the renormalizable
couplings on our brane must be baryon- and lepton-number
conserving.

\subsection{How Large Are the Extra Dimensions?}

The predictiveness of the SLED proposal lies in the inability to
make the extra dimensions smaller --- and so easier to hide ---
without destroying the explanation of the observed Dark Energy. In
this section we discuss the most restrictive observational limit
on the possible size the extra dimensions can take, in order to
show that this is (just) large enough to be consistent with the
size of the Dark Energy density. Our purpose is to argue that this
consistency does not allow much freedom to shrink the extra
dimensions, and so fairly robustly sets the scale of the extra
dimensions, as well as constraining the kind of 6D supergravities
which can be entertained.

\nisubsubsection{Supernova Bounds}

\noindent The strongest limit on the size of two large extra
dimensions comes from the constraint that energy loss into Kaluza
Klein modes not provide too efficient an energy-loss mechanism for
supernovae \cite{LEDastrobounds,HR}. This process has been studied
in detail for the special case of the radiation of gravitons into
the bulk, with the recent study \cite{HR} showing that an
acceptably small energy-loss rate requires the 6D gravitational
scale, $M_g$, to satisfy $M_g > 8.9$ TeV.\footnote{Our conventions
define the 4D Planck scale by $M_p^{2} = (8 \pi G_N)^{-1} = M_g^4
r^2$ when the two extra dimensions have volume $V = r^2$.} For
unwarped extra dimensions this requires the extra-dimensional size
to be $r < 10 \; \mu$m.\footnote{Ref.~\cite{HR} quotes $r < 1.6 \;
\mu$m, but uses conventions where the extra-dimensional volume is
given by $V = (2 \pi r)^2$.}

This limit may be even stronger for SLED, because the theory in
the bulk is supersymmetric and so can offer more modes into which
energy may be lost. Although the precise loss rate is not yet
computed for SLED, a simple estimate is obtained by scaling the
graviton energy loss rate by the total number of degrees of
freedom, ${\cal N}$, in the extra dimensions whose couplings allow
them to be emitted singly starting only from brane states. For
instance if only gravitons could be emitted, this number would be
${\cal N}_{\rm LED} = 3$ for particle collisions in vacuo on flat
3-branes, since this corresponds to the 3 of the 9 6D graviton
polarizations which can couple to purely 4D stress energy. We
therefore estimate the loss rate into SLED bulk states to be
\beq
    \Gamma_{SLED} \approx \Gamma_{LED} \left( \frac{{\cal N}}{3}
    \right) \,.
\eeq
where $\Gamma_{LED} \propto M_g^{-4}$ is the standard result for
graviton emission by flat branes. This leads the SLED bound on
$M_g$ to become
\beq
    M_g > \left( \frac{{\cal N}}{3} \right)^{1/4}
    8.9 \;\hbox{TeV} \, .
\eeq

The bound therefore depends on the number ${\cal N}$, although the
$1/4$ power in the last expression shows that this dependence is
not inordinately strong. This number is clearly model-dependent in
two separate ways. First, it depends on the total number of states
in the bulk which are massless in the 6D sense, and so which are
potentially available channels for carrying off energy. This
depends on the multiplet content of the supergravity under
consideration, and can be quite large for the chiral theories. In
the same counting that assigns 9 spin states to the 6D graviton,
the (massless) 6D gravity, gauge and matter supermultiplets
respectively have $32 = 16_F + 16_B$, $16 = 8_F + 8_B$ and $8 =
4_F + 4_B$ spin states, where $B$ and $F$ denote whether the
states being counted are bosons or fermions. The number of
massless 6D bulk states can therefore be as small as 32 for
non-chiral 6D supergravities \cite{Ungauged,Romans}, with no
matter multiplets in the bulk. Alternatively, for chiral, gauged
6D \cite{NS} supergravity in the bulk, anomaly cancellation
requires the number of matter and gauge multiplets to satisfy the
condition $N_m = N_g + 244$ \cite{6DAC},\footnote{More precisely,
the anomaly-cancellation condition is $N_m - N_g = 273 - 29 N_t$,
where $N_t$ is the number of tensor multiplets. Although the
option $N_t = 1$ allows the writing of a Lorentz-invariant action,
$N_t > 1$ could permit anomaly cancellation with fewer matter
fields.} and so even choosing $N_g = 0$ in this case implies there
must be 1952 massless 6D spin states in the bulk. If all of these
contributed as strongly to the energy-loss as an allowed graviton
mode, this would lead to the bound $M_g > 34$ TeV and so $r < 0.67
\; \mu$m. Fortunately, as we now argue, the real bound is likely
to be less restrictive than this, even for the chiral theories.

The real bounds are likely to be weaker because the second source
of model-dependence acts to reduce ${\cal N}$, and so to weaken
the bounds on $M_g$ and $r$.\footnote{We thank the referee for
stressing that not all bulk particles are likely to be emitted as
strongly as are some graviton modes.} This second model-dependence
is to do with the size of the amplitude for emitting a single bulk
particle given only brane-bound particles in the initial state.
This amplitude depends on both the kind of bulk particle involved,
and on the kinds of effective interactions with the bulk which
arise on the branes. For instance, even for gravitons not all of
the 9 possible 6D spin states are emitted from particles moving
along a flat brane because not all of the graviton polarizations
couple to the purely 4D components of the matter stress tensor.
More graviton spins would couple if the brane were bent or moving
in the transverse dimensions, or in the presence of more
complicated internal initial states on the brane.

Similar statements apply for the other bulk fields, whose
couplings also depend on the kinds of interactions which are
present between brane and bulk particles. For example, the fact
that supersymmetry is broken on our brane implies the existence of
model-independent couplings between the brane-based Goldstone
fermion and the bulk gravitino \cite{nonlinearbranesugra}. For
other bulk particles, preliminary examination of the simplest
kinds of brane couplings to 6D scalars and vectors is given in
refs.~\cite{LEDSusyCollider} and \cite{AbelSchofield}, and similar
estimates for the fermion couplings relevant to neutrino
interactions are given below. From these it is clear that the
strengths of these couplings --- and so also the strengths of the
resulting bounds --- can vary considerably, depending on the brane
details.

The weakest possible bound on $r$ occurs for models for which only
the graviton and gravitino couple linearly to brane matter.
Because of the weak dependence on ${\cal N}$ in $\Gamma_{SLED}$
this leads to constraints which are essentially the same as for
pure-graviton emission in the nonsupersymmetric case.
Alternatively, if ${\cal N} \sim 100$ then the LED constraint
would be marginally strengthened to $M_g \gsim 12$ TeV and $r
\lsim 5.3 \;\mu$m. It is clear that these are only rough estimates
and that an explicit calculation of the rate of energy lost is
desirable for specific models (but lies beyond the scope of the
present article).

There are also other, nominally stronger, bounds on
extra-dimensional models which come from the non-observance of
Kaluza-Klein modes decaying into photons after having been
produced in supernovae or in the early universe. We ignore these
bounds for the present purposes, since unlike the bound just
discussed they can be completely evaded depending on the details
of the model. For instance, they do not arise if KK modes can
efficiently decay into invisible light modes on other branes. We
regard the model building which such an evasion requires to be
well worth the cost if the resulting theory can make progress on
the much more difficult cosmological constant problem.

\nisubsubsection{The Dark Energy Constraint}

\noindent Balanced against these bounds is the requirement of a
sufficiently small Dark Energy density. Within SLED models this
energy arises as a Casimir energy, whose evaluation (including the
back-reaction of the branes on the bulk geometry) has not yet been
done, but is in progress. Perhaps the most interesting thing about
the SLED proposal is that within it the success of the description
of the size of the cosmological constant comes down to the
discussion of the $O(1)$ factors, which we describe here.

In the absence of a definitive calculation of the Casimir energy
in the presence of branes within a compact two-dimensional
solution to 6D supergravity, a reasonable picture of the $r$
dependence of the Casimir energy can be obtained by dimensional
analysis supplemented by the tracking of the large logarithms
which crop up due to the ultraviolet sensitivity of the result to
large mass scales (like $M_g$) \cite{Doug}. Taking care to write
the result in the Einstein frame (for which Newton's constant is
not field dependent), the result is
\beq
    V(r) = \frac{A}{r^4} \left[1 - a \, \log(M_p r) + \frac{b}{2} \,
    \log^2(M_p r) + \cdots
    \right] + \cdots \,,
\eeq
where $A$, $a$ and $b$ are dimensionless constants, the first
ellipses represent possible terms involving higher powers of $\log
\, r$, and the second ellipses indicate terms involving higher
powers of $1/r$. As written, $V$ is positive for all $r$ provided
$A > 0$ and $2b > a^2$.

Besides being the likely result of explicit calculations, this
potential is known to be able to describe a phenomenologically
acceptable description of time-dependent Dark Energy
\cite{AS,ABRS2}. If $a^2 + (b/4)^2 - 2b > 0$ the potential has a
local minimum at $r_-$ and a maximum at $r_+$, where $4b \log(M_p
r_\pm) = b + 4a \pm \Delta$, and $\Delta = 4[a^2 + (b/4)^2 - 2
b]^{1/2}$. $r_+$ and $r_-$ are naturally of order 10 $\mu$m
provided $a/b \approx 70$. It should be noticed that the success
of the cosmologies of refs.~\cite{AS,ABRS2} do not rely on the
existence of a minimum or maximum, but in them the acceleration of
the present-day Dark Energy is described by the epoch when $r$ is
in the vicinity of $r_\pm$, at which point we have, for instance
\beq
    \lambda \approx V(r_+) = \frac{A}{16 r^4}\Bigl[ b + \Delta \Bigr] \,.
\eeq
Given estimates for $A$, $a$ and $b$ we may determine what the
smallest value is that $r$ can take at present without having
$\lambda$ be unacceptably large.

The constant $A$ is obtained from a Casimir energy calculation,
and depends on details of the extra-dimensional geometry and the
boundary conditions satisfied by the various bulk fields which are
massless in 6 dimensions. For the present purposes we estimate the
size of $A$ using calculations of the Casimir energy on a 2-torus,
with supersymmetry broken by choosing different boundary
conditions for bosons and fermions
\cite{ScherkSchwarz}.\footnote{This roughly captures what is
expected from the back reaction of the branes, since these
introduce conical singularities into the geometry which affect the
boundary conditions of different spins in different ways.}
Calculations such as these have been computed in many places for
torii. For torii $A$ depends on the various toroidal shape moduli
\cite{Poppitz} but for simplicity we ignore these moduli and
estimate $A$ using the numerical results of ref.~\cite{ABRS1} for
a square torus. In this case $A = - 0.15$ for a periodic massless
scalar and $A = 0.23$ for a completely antiperiodic massless
spin-half fermion. This leads to the estimate for $A$ for a
supersymmetric theory, with supersymmetry broken by boundary
conditions, of $A \approx (0.23 - 0.15) N = 0.08 N$, where $N$
counts the number of bosonic degrees of freedom in the bulk ({\it
e.g.} $N = 16 + 8N_g + 4 N_m$). With $N = 16$ we have $A \approx
1.3$.

The constants $a$ and $b$ are likely to arise from logarithmic
divergences in 6 dimensions, and although they are not yet
calculated we expect $b \sim a^2$, with $a \sim N/(2 \pi)^3$, for
$N$ an indicator of the number of degrees of freedom in the bulk
(and so $a/b \sim 70$ is possible if $N \sim 10$, as for the 6D
supergravity multiplet). Taking $\Delta \approx O(a)$ we then find
\beq \label{lambdaestimate}
    \lambda \sim \frac{Aa}{16 r^4} \sim \frac{1.3}{70 \times 16 r^4}
    \sim \frac{1}{(6 r)^4} \, .
\eeq

This estimate is this section's main result. Given that (10
$\mu$m)${}^{-1} = 0.020$ eV, we see that $\lambda$ takes the
observed value $(0.003$ eV)${}^4$ when $r$ is close to its current
observational upper limit, $r \sim 5$ $\mu$m (up to within the
uncertainties of the estimates given). Clearly these estimates
strongly motivate a more detailed analysis of the Casimir energy
within specific SLED vacua.

\subsection{More Fundamental Origins?}

One of the principal motivations for the supersymmetric Standard
Model comes from the plausibility of its arising as the low-energy
limit of a well-motivated, more fundamental, theory at higher
energies. It is worth asking to what extent the same might also be
true for the SLED proposal. Since it is difficult to say anything
concrete about this without having some framework in mind for what
the kind of fundamental physics might be, we choose to cast this
question in terms of what is at present probably our
best-motivated fundamental supersymmetric theory: string theory.
Since SLED is a particular case of the brane-world scenario, it is
plausible that it could be obtained from string theory since this
naturally includes D-branes and similar objects onto which it is
known that Standard Model particles can be trapped.

A successful embedding of SLED into string theory consists of
identifying string vacua for which the low-energy excitations
describe both the 6D supergravity of the bulk and the degrees of
freedom trapped on the branes. Indeed, the non-chiral (gauged and
ungauged) 6D supergravities have a known string provenance in this
way as dimensional reductions of 10D string vacua, and although a
complete string derivation of the chiral 6D theory is not yet
known --- see however \cite{Pedigree} --- it is plausibly
derivable as the low-energy limit of 10D heterotic string theory
compactified on $K_3$.

Knowing how the branes arise within string theory is also
important, and this has three different aspects. It involves
finding string vacua having 3-branes in the effective low-energy
6D supergravity; it involves finding the particles of the Standard
Model living on one of them; and it involves obtaining the
brane-bulk couplings which are required by the mechanism of making
the observed dark energy small \cite{Towards,Warped}.\footnote{The
precise requirement is for the dilaton coupling to vanish in the
6D Einstein frame.} A complete construction has not yet been
accomplished, although branes having the required dimension and
dilaton couplings may be obtainable either from D3 branes in 10D
Type IIB string vacua, or from NS5-branes wrapped on 2-cycles in
10D heterotic (or 11D $M$-Theory) compactifications to 6
dimensions.

An important difficulty in obtaining MSLED models from string
theory is also familiar from attempts to find the MSSM within
string theory: there are typically too many degrees of freedom in
the low-energy theory. This is a particularly pointed problem for
MSLED given the strong constraints against the existence of too
many light modes in the bulk. Furthermore, because the
supersymmetry-breaking scale in the bulk is so low, MSLED models
cannot blindly appeal to supersymmetry breaking to give these
moduli TeV-scale masses. What is required is a compactification to
6 dimensions for which the moduli for the internal dimensions (and
any other light fields) are fixed at the string scale without
breaking the bulk 6D supersymmetry. These conditions make a string
theory derivation of the MSLED challenging, but perhaps not
necessarily more so than a similar derivation of the MSSM.

It is this problem of ubiquitous light fields which argues against
the simplest compactifications of 10D models to 6D. For example,
for the perturbative heterotic string (or in the 11D Horava-Witten
scenario) all of the gauge degrees of freedom coming from the
$E_8\times E_8$ or $SO(32)$ gauge sectors are expected to populate
the bulk in a 6D compactification. Fewer such fields may arise if
the low-energy gauge sector lives at the singularities of
manifolds of $G_2$ holonomy, but these are not sufficiently well
understood at the moment to allow the construction of explicit
models.

Better prospects for finding SLED vacua with acceptably few light
fields in the bulk
may come from Type II theories (and their orientifolds
--- including Type I vacua) with the gauge sector localized on
D-branes. In general,
for the quasi-realistic examples found to date the
Standard Model fields are localized on a D3 brane which is located
at singularities in the bulk, or they are localized on the
intersection of wrapped D6 or D5 branes (see for instance
\cite{real1,real2}). Furthermore,
 for Type IIB theories it is known that
compactifications with  fluxes can fix many of the
low-energy moduli which would have arisen without the
fluxes \cite{GKP,kklt}. These kinds of models also have the advantage
that quasi-realistic models have been constructed for which the
Standard Model (or its supersymmetric extension) is known to live
on a brane \cite{throat}.

The 10D scenarios of interest here require a compactification for which 4
of the internal 6 dimensions are much smaller than the remaining
2. How much smaller they are is likely to depend on the details of
the compactification, and in particular on whether any of the 4
smaller dimensions should be warped. For instance, the
compactification might be envisioned to arise simply as an
orientifold of the product space $K_3 \times T^2$ or in a more
complicated way as an elliptically-fibred Calabi-Yau
compactification --- {\it i.e.} a compactification which is
locally $T^2\times B$ --- for which the elliptic fiber is a $T^2$
whose size varies with position in $B$.

If the small 4 dimensions are unwarped and have volume $V_4 =
\ell^4$, then the effective 6D gravity scale is related to the
string scale, $M_s$ and $\ell$ by $M_g = e^{-\Phi/2} M_s^2 \ell$,
where $\Phi$ is the 10D dilaton and we assume a string-frame 10D
Einstein action of the usual form ${\cal L}_{10} \propto
e^{-2\Phi} R$. Within a perturbative low-energy framework with no
particularly small numbers put in by hand ({\it i.e.} for which
$e^\Phi \lsim 1$ and $M_s \ell \gsim 1$) we are led to a picture
where both the string scale and the size of the 4 small dimensions
are not so different from the 6D gravity scale $M_g$. In this
unwarped scenario we therefore expect to find TeV-scale phenomena
associated with the small 4 extra dimensions and with string
physics itself. Notice also that the conditions for the validity
of perturbation theory, $e^\Phi \ll 1$ and $M_s \ell \gg 1$, both
imply $M_g > M_s$, and so within the unwarped picture string
physics can be expected to appear at energies {\it lower} than
$M_g \sim 10$ TeV.

Alternatively, if the small 4 dimensions are strongly warped we
may have $\ell^{-1} \sim M_s \gg M_g$ with the hierarchy between
these two scales arising from the warping within the four extra
dimensions. Scenarios in which the string scale is still close to
the Planck scale or  the intermediate scale $M_s\sim 10^{11}$ GeV
\cite{intermediate} could be possible in this setting. Within this
kind of hybrid Randall-Sundrum \cite{RS}/ADD scenario we need not
expect to find more string or extra-dimensional physics at the TeV
scale beyond that expected purely from the 6D degrees of freedom,
but it may instead have interesting cosmological implications.

An interesting remark in this regard for both of these kinds of
pictures is that the desired string compactification need not
stabilize the large two extra dimensions, since these can instead
be still dynamically evolving at present along the lines described
in previous sections. All that is required is that the
loop-generated potential for the 2D moduli have a form similar to
that described in earlier sections. In particular, it would be
interesting to be able to predict the initial features of this
late-time cosmology as the final state of an earlier inflationary
period, given the recent progress towards finding inflation within
string theory \cite{kklt,StringInflation,inflation2}.

\section{Implications for Particle Physics}

In this section we briefly summarize the main implications which
the MSLED proposal can be expected to have for current particle
physics experiments. We expect these to come in two broad classes
of phenomena: neutrino physics and collider experiments at TeV
scales. These two kinds of observables differ in the extent to
which they test the general SLED proposal or its more specific
MSLED variant. In this section we argue that SLED models very
generically predict new effects for TeV colliders. We also explore
whether the MSLED proposal predicts new neutrino physics. In
passing we remark that that the minimal MSLED predicts only
unobservably small new phenomena for flavor-changing experiments
such as those being done with $K$ and $B$ mesons.

\subsection{Neutrino Physics}

The SLED proposal relies on the numerical coincidence between the
value of $1/r$ which is required in 6 (unwarped) dimensions by the
explanation of the Dark Energy density, and by the explanation of
the size of the electro-weak/gravitational hierarchy, $M_g^2/M_p$.
Since the value for $1/r$ which results is also close to the mass
differences, $(\Delta m^2_{atm})^{1/2} = 0.05$ eV, appearing in
atmospheric neutrino oscillations \cite{AtmosNu} --- or $(\Delta
m^2_\odot)^{1/2} = 0.007$ eV for solar neutrinos \cite{SolarNu}
--- it is natural to ask whether there might be a physical reason
for this.

\nisubsubsection{Models Without Brane-Bulk Neutrino Mixing}

\noindent It must be emphasized at the outset that in general
there {\it need} not be a connection between neutrino masses and
the extra-dimensional scales within SLED. For instance, we might
imagine stepping outside of MSLED and accounting for neutrino
oscillations by simply putting 3 light sterile neutrinos onto our
brane along with the Standard Model, with their masses tuned to
the appropriate values by hand. Alternatively, even within MSLED
we could obtain neutrino masses without introducing new low-energy
brane degrees of freedom by contemplating the usual dimension-5
interaction on our brane,
\beq
    {\cal L}_5 \propto \frac{h^2}{M_g} \, (L^i \gamma_L L^j) H_i H_j
    + \hbox{c.c.} \,,
\eeq
with $L = \left( {\nu \atop e} \right)$ denoting the Standard
Model leptons, and $H$ representing the Higgs doublet. The
challenge in this kind of explanation of neutrino masses would be
to understand why the dimensionless coupling $h$ should be so
extremely small: $h^2 \sim 10^{-11}$. Such small couplings are
natural in the sense that they are stable under renormalization,
and their explanation would require an understanding of why the
underlying higher-dimensional physics (such as the appropriate
vacuum of string theory) preserves baryon and lepton number to
high accuracy. In either of these two ways of understanding
neutrino masses, it is difficult to make a definitive prediction
for what to expect in neutrino experiments from SLED or MSLED.

\nisubsubsection{Models With Brane-Bulk Neutrino Mixing}

\noindent Things would be more interesting, however, if there
should be a nontrivial mixing between brane-bound neutrinos and
bulk fermions. This could arise through an interaction of the form
\beq \label{bulkbranenumixing}
    S_{{\rm int},N} = \int d^4x \; \left[ {g_{aI}}
    (L^i_a \gamma_L N^I) H_i + \hbox{c.c.} \right] \, ,
\eeq
where $N^I$ denotes a collection of 6D bulk fermions and $g_{aI}$
are coupling constants having dimensions of inverse mass, which we
expect to be roughly of order $M_g^{-1}$ in size. In this
expression the index $i = 1,2$ is an $SU_L(2)$ gauge index while
the index $a = 1,2,3$ labels fermion generations.

Models of this class are studied in detail in
ref.~\cite{ExtraDimNus,Cirelli,BBLED}. What has poisoned attempts
to build 6D neutrino phenomenology along these lines is the
efficiency with which active-sterile neutrino mixing can drain
energy away from astrophysical sources like supernovae. This can
happen in either of two separate ways: either from losses due to
incoherent radiation of bulk neutrinos; or from losses due to
resonant oscillations into bulk neutrinos. Because of these
constraints previous workers typically choose their models to be
effectively 5-dimensional --- an option which seems unavailable to
us in the SLED context.

It is the first of these which seems the most dangerous in the
present instance, since the constraint relies simply on the
emission rate into sterile modes being of order $\Gamma_s \approx
(m/E)^2 \Gamma_\nu$, where $\Gamma_\nu$ is the rate for emitting
the standard active neutrinos, $E \sim 100$ MeV is the typical
neutrino energy, and $m \sim g_{aI} v/r$ is the mass term which is
responsible for the brane-bulk neutrino mixing. Because of the
enormous number of bulk states which can be radiated, the energy
loss due to this emission channel is too large provided $m r \sim
g_{aI}v \lsim 2 \pi \times 10^{-4}$. At face value such small
values are incompatible with the regime $g_{aI}v \sim O(1)$ which
are required by neutrino phenomenology.

For this reason we do not further pursue brane-bulk neutrino
mixing within SLED, although one always wonders whether a
constraint which relies in this way on supernova physics might
have loopholes to do with our fairly poor understanding of
neutrino physics within dense supernova cores. If
ref.~\cite{Cirelli} is correct, this kind of worry actually may be
justified for the bound on brane-bulk neutrino mixing which is
derived from {\it coherent} bulk neutrino production within
supernovae. At face value this bound can be even stronger than the
one just discussed, and so potentially is even more dangerous.
However, ref.~\cite{Cirelli} claims that this bound is weaker than
had been previously thought, because of a feedback effect due to
neutrinos modifying their own environment within supernovae in
such a way that the beginnings of efficient oscillation into bulk
modes acts automatically to switch off the bulk radiation. If so,
then this second bound need not be a concern for MSLED models. We
believe it to be worth exploring in more detail the extent to
which these bounds rule out nontrivial bulk-brane neutrino mixing
within the MSLED framework.

\subsection{Signatures at Colliders}

A robust consequence of the SLED proposal is the existence of many
bulk Kaluza-Klein modes whose mass spacing is of order $2\pi/r
\sim 0.3$ eV. Although each of these modes is coupled to ordinary
particles only with gravitational strength, their enormous phase
space at TeV energies makes their collective effects enter into
observables at collider experiments with rates of order
\beq
    \Gamma(E) \propto \left( \frac{1}{M^2_p} \right) (E r)^2 \sim
    \left( \frac{E^2}{M_g^4} \right) \,,
\eeq
whose size is controlled by powers of $M_g$ rather than $M_p$. For
this reason the production of missing energy at TeV scale
colliders is a robust signal of the SLED proposal.

Of course, the same arguments apply equally well to the
non-supersymmetric LED proposal, where they again argue for a
robust collider signal, and this has led to extensive studies of
the possible phenomenological signatures of graviton emission into
the bulk \cite{LEDColliderTh,LEDColliderEx}. Since these studies
typically show that observable effects require $M_g \lsim 1$ TeV,
one might worry that the supernova constraint $M_g \sim 10$ TeV
must preclude the expected SLED collider signal from being
observably large.

We argue that this conclusion may be too pessimistic for several
reasons. First, these calculations were done only for the graviton
and we have already seen that radiation rates into the bulk are
more efficient in SLED because of the existence there of all of
the graviton's super-partners \cite{LEDSusyCollider}. Furthermore,
the pessimistic conclusion also relies too heavily on there being
no new physics at energies which are lower than the scale $M_g$,
even though $M_g$ is really only the scale of 6D gravity as
measured by the higher-dimensional Newton constant. As such, it
need not be the threshold at which new physics first emerges. (A
similar error would be made for the weak interactions if the Fermi
constant, $G_F^{-1/2} \sim 300$ GeV, were used to infer where the
scale where the physics of the electro-weak interactions first
appears. In reality we know that this physics starts at $M_w = 80$
GeV rather than 300 GeV, because of the appearance of small
dimensionless couplings in the relation between $G_F$ and $M_w$.)
In the same fashion, it is likely that within the SLED proposal
the string scale is at or below the scale $M_g$, making some
states potentially available to experiment at scales of a few TeV.

Furthermore, depending on how it arises within a more microscopic
theory like string theory, there is a possibility in the SLED
picture that there are also Kaluza-Klein and winding states
associated with the `other' 4 compact dimensions obtained when
compactifying from 10 to 6 dimensions. As we saw in earlier
sections, these should also have masses in the TeV regime provided
the 4 small internal dimensions are not strongly warped. In this
regime collider reactions should resemble string collisions near
Planckian energies \cite{StringsPlanck}. A more precise study of
the phenomenology of these modes requires the study of the string
compactifications from 10 to 6 dimensions which lead to the kinds
of 6D models of present interest.

The most likely collider signal (besides the direct production of
new string or KK states) of SLED physics is therefore likely to be
missing energy, which can be produced in association with an
isolated jet or lepton. Such a signal is unlikely to be confused
with the missing energy signals of alternative proposals, such as
the Minimal Supersymmetric Standard Model (MSSM). There are also
discussions in the literature \cite{BHatLHC,giota} about the
possibility of producing mini black holes when collider energies
approach the 6D gravitational scale. However, these calculations
are still subject to considerable theoretical uncertainty. Among
the open and speculative issues are: the mass at which the black
hole description starts being valid \cite{Giddings}; the
applicability of using Thorne's hoop conjecture when computing the
black hole production cross-section \cite{Thorne}; the use of the
generalized uncertainty principle (which the authors of
ref.~\cite{gup} claim implies a large increase in the energy
necessary to form a black hole, putting them beyond the reach of
the LHC); and the validity of the description of black-hole
creation in terms of the collision of Aichelburg-Sexl shock waves
\cite{aichel}.

\subsection{Flavour Physics}

Another way to distinguish the predictions of MSLED from alternate
proposals is in their implications for flavor-changing physics as
seen in $B$ and $K$ factories. Unlike theories like the MSSM,
MSLED does not predict new phenomena in these experiments because
its flavor physics is essentially described by the Standard Model.
Indeed, a virtue of the MSLED proposal is that it retains many of
the Standard Model successes in this regard, such as the natural
understanding --- through the GIM mechanism --- of why
flavour-changing neutral currents are small, and of why CP
violation is small for $K$ mesons physics even though the
intrinsic CP-violating phase in the Cabbibo-Kobayashi-Maskawa
matrix is not small. At present (with very few, controversial
exceptions) the existing data on weak decays of hadrons, including
rare decays ($B\to X_s \gamma$ \cite{bsg}, ...) and CP violating
observables (CP asymmetry of $B\to J/\Psi K_S$ \cite{bjpsiks},...)
is described within theoretical and experimental uncertainties by
the SM, and so MSLED inherits this success. The CKM mechanism has
passed its first precision tests successfully and is very likely
the dominant source of CP violation in flavour-changing processes.

The only hope to see new effects in flavour-changing meson
experiments within the MSLED framework would be if flavor-diagonal
new physics taken together with the ordinary weak interactions
could be separated from penguin contributions to various meson
decays. Unfortunately, any such an analysis is likely to require a
much more accurate understanding of penguin processes and matrix
elements than is currently feasible.

It must be emphasized that these statements hold only for the
minimal MSLED proposal, and do not follow robustly for all models
in the SLED class. This is because one may always use the freedom
to introduce new flavour physics by complicating the physics of
our brane, without ruining the success of the basic 6D SLED
mechanism.

\section{Cosmology}

Since the main motivation of the SLED proposal comes from
cosmology, it is not surprising that some of its observational
implications are cosmological. In this section we briefly
summarize the opportunities for testing the proposal which can be
explored over comparatively long distance scales. These come in
three different types: implications for Dark Energy, implications
for tests of General Relativity on large and small distance
scales, and implications for Dark Matter. We close with some
speculations about how inflation might be embedded into the SLED
picture.

\subsection{Dark Energy}

Besides providing a natural size for the Dark Energy density, the
SLED proposal makes many further predictions concerning the
properties of Dark Energy. First and foremost, it predicts that
the Dark Energy is dynamically evolving in time even now, and so
is not simply a cosmological constant. Rather it is what has come
to be known as a `quintessence' model \cite{Quintessence},
involving a cosmologically evolving 4D scalar field (or fields)
whose microscopic origin is the overall breathing mode of the 2
large extra dimensions, $r$ (plus possibly various shape moduli
for these same dimensions).

Besides predicting the Dark Energy to be cosmologically evolving,
SLED also predicts a very specific form for the quintessence
field's scalar potential. Given that the Einstein-Hilbert action
implies the radion kinetic energy has the form $M_p^2
(\partial_\mu r \, \partial^\mu r)/r^2$, the
canonically-normalized field is $\varphi = M_p \log(M_p r)$,
leading to an exponential potential with a power-law prefactor.
This is a form originally proposed by ref.~\cite{AS}.\footnote{The
potential for any shape moduli depends on more detailed
information about the geometry of the two large dimensions.}

Because the radion's scalar potential has a previously-discussed
form, it follows that the SLED predictions for Dark Energy
evolution are likely to resemble those of refs.~\cite{AS,ABRS2}.
This shows that SLED can share the viable phenomenology of these
models, as well as their theoretical puzzles. In particular,
although these models have many tracker solutions \cite{ABRS2},
their successful description of present-day cosmology does not
rely on them and instead depends on the initial conditions of the
radion field in an important way. This in turn implies that a
completely adequate description of the cosmology of Dark Energy is
likely to require a better understanding of these initial
conditions, possibly as a consequence of an earlier inflationary
epoch \cite{QuintInflation}. (Of course, these conclusions must be
re-examined to the extent that shape moduli are also
cosmologically active --- see for instance \cite{Poppitz} for a
preliminary examination of this issue for torii.)

It should be stressed that these predictions for the Dark Energy
are robust consequences of the relaxation mechanism with which the
SLED proposal cancels the contributions of the brane tensions to
the Dark Energy, since this mechanism implies the existence of a
very light scalar field, $r$, whose mass is of order the
present-day Hubble scale, $m_\varphi \sim H_0 \sim 10^{-33}$ eV.

\nisubsubsection{Quintessential Naturalness Issues}

\noindent The existence of a scalar as light as $m_\varphi \sim
10^{-33}$ eV brings with it a further naturalness issue, which
asks how such a small scalar mass can be stable against radiative
corrections as physics associated with energy scales larger than
$m_\varphi$ are integrated out \cite{QuintNat}. Remarkably, this
naturalness problem is automatically explained in the SLED
proposal as a consequence of the natural explanation of the small
size of the observed Dark Energy density. This SLED explanation
proceeds essentially along the lines anticipated in
refs.~\cite{ABRS1,ABRS2}, and proceeds as follows.

The explanation has two parts, depending on the energies being
integrated out. For particles with masses, $m$, lying between
$m_\varphi \sim 10^{-33}$ eV and $1/r \sim 0.01$ eV, the effective
theory is 4-dimensional, and standard arguments apply. These state
that the integration over modes having mass $m$ which couple to
the quintessence field with strength $g$ should contribute (at one
loop) an amount $\delta m_\varphi \sim mg/(4 \pi)$. Since in the
present case the quintessence field couples with gravitational
strength, the relevant coupling is of order $g \sim m/M_p$,
leading to the estimate
\beq
    \delta m_\varphi \sim \frac{1}{4 \pi} \, \left( \frac{m^2}{M_p}
    \right) \, ,
\eeq
which is acceptably large because this calculation necessarily
presupposed $m \lsim 1/r \sim 0.01$ eV in order to be performed in
4 dimensions.

For particles more massive than $1/r$ the analysis must be done in
6 dimensions, and in this case the stability of $m_\varphi$ is
ensured by the same mechanism which keeps the Dark Energy density
sufficiently small. That is, to the extent that the radion
potential is of order $V(r) \sim 1/r^4$, the radion mass is
naturally $m_\varphi \sim 1/(M_p r^2)$ and so has the right size
for $r \sim 10$ $\mu$m. The SLED proposal itself explains why
particles on the brane do not contribute to $V(r)$ at all
regardless of their masses. On the other hand, for particles in
the bulk the integration over modes of mass $m$ can generate
dangerous terms which are of order $\delta V \sim m^2/r^2$. If
generated, these would both give too large a vacuum energy
density, and contribute the too-large amount $\delta m_\varphi
\sim m/(M_gr)$.\footnote{Although quoted above in Jordan-frame
units, these expressions for $m_\varphi$ are most easily
understood in the 4D Einstein frame using the
canonically-normalized field $\varphi$, with $M_p r \sim
e^{\varphi/M_p}$. In terms of this we have $V(r) \sim m^2/r^2$
implying $V(\varphi) \sim (m^2 M_p^4/M_g^2) \, e^{-2\varphi/M_p}$
while $V(r) \sim 1/r^4$ implies $V(\varphi) \sim M_p^4 \, e^{-4
\varphi/M_p}$.} Notice that the result $m_\varphi \sim 1/r$
obtained when $m \sim M_g$ corresponds to the usual estimate for
the loop-induced radion mass in non-supersymmetric LED models.

Typically, the absence of these dangerous $m^2/r^2$ terms need
only be ensured at one loop in the bulk theory, and they are in
particular absent in the 6D theories of most practical interest
(such as those obtained by compactification from 10D supergravity)
\cite{Update,Doug}.

\subsection{Tests of Gravity}

There are two types of changes to gravitational physics which must
follow in any variant of the SLED proposal (and not just for its
MSLED version). These involve tests of Newton's inverse-square law
on distances of order $m_{KK}^{-1} \sim r/(2\pi)$, as well as
very-long-distance modifications to gravity over scales up to the
present-day Hubble length, $H_0^{-1}$.

\nisubsubsection{Short-Distance Tests}

The deviations from Newton's inverse-square law for gravitation
must break down in SLED for the same reason as it does within the
earlier LED proposal: the effects of ordinary graviton exchange
begin to compete at these distances with the exchange of various
Kaluza-Klein modes from the two large extra dimensions. The range
over which the deviations from the inverse-square law arise is set
by the lightest of the nonzero masses of the Kaluza-Klein modes.
Although the precise value of this lightest mass depends on the
details of the precise shape of the extra dimensions, we have seen
above that they are of order $m_{KK} \sim 2\pi c/r$ with $c = 1$
for a toroidal compactification. It would be worthwhile computing
the form of the predicted force law in more detail for
representative geometries, given the many bulk fields in SLED
whose Kaluza Klein modes could be relevant.

The bad news here is that the requirement $r \sim 5$ $\mu$m, which
follows from the Dark Energy density, implies $r/(2 \pi) \sim 0.8$
$\mu$m, which is some two orders of magnitude below the 100 $\mu$m
range to which the best present searches for these effects are
sensitive \cite{NLTests,NLTestsRev}. They might not be beyond the
reach of future tests based on precision measurements of the
Casimir effect \cite{NLTestsRev}, however. The good news is that
there absolutely must be an effect, and that the $O(1 \; \mu$m)
range cannot be moved to still smaller values to evade better
tests, once these become available.

\nisubsubsection{Long-Distance Tests}

SLED models must also predict very long-range forces of roughly
gravitational strength, of the general scalar-tensor form. This
follows quite generally as a consequence of the relaxation
mechanism it predicts for the Dark Energy density, which we have
seen leads naturally to one or more fields, $\varphi$, whose mass
is presently incredibly small, $m_\varphi \sim H_0 \sim 10^{-33}$
eV. Unlike almost all other quintessence proposals, we have seen
how this small scalar mass can be naturally stable against quantum
effects within the SLED framework.

Such a light scalar is strongly constrained by searches for a
very-long-range force which competes with gravity \cite{LRTests}.
At the classical level the theory is very predictive because the
breathing mode of an extra dimension is known to couple to
ordinary matter on our brane through their contributions to the
trace of the stress tensor, $T^\mu_\mu$, and at first sight this
seems to be a fatal prediction since gravitational-strength
couplings of this form can already be ruled out with some
precision.

A more detailed look is again more interesting, and whether these
tests falsify SLED cosmology relies on more detailed calculations
of the $r$-field couplings than are presently available. That they
do can be seen from models such as those of
refs.~\cite{ABRS1,ABRS2}, for which the precise strength and
coupling of the light field can actually depend on some of the
details of its cosmological evolution, since they can evolve
during the history of the universe. They can do so because in the
models investigated these couplings are field-dependent. Since the
measurements which constrain these models are performed only
during the present epoch, they are satisfied provided the
couplings happen to be small at present, and this is what happens
in the cosmological models of \cite{ABRS2}. (It has also been
claimed that an evolution to small couplings is an attractor in
the space of solutions of related scalar-tensor models
\cite{Damour}.)

Similar effects can arise within SLED scenarios because loop
effects can introduce an $r$-dependence into otherwise
field-independent quantities. Clearly the question of how big
these couplings now are (and whether their motion can be traced
within the domain of perturbation theory) can be addressed in a
more focussed way given a real calculation of the potential and
interactions of the field $\varphi$ (and any other light fields)
from a specific microscopic model for the extra-dimensional
geometry. This provides yet more motivation for exploring the
details of such compactifications, including loop effects, more
closely.

The possibility that there can be very light scalar fields around
providing us with interesting non-standard gravitational physics
on the longest distance scales is particularly exciting given the
recent opportunities for testing this kind of scenario. It
motivates, in particular, a more careful exploration of the
phenomenology of scalar tensor theories for solar-system tests, as
well as for tests in more exotic settings (like the
recently-discovered system consisting of two pulsars which orbit
one another \cite{HolyGrail}). Given a theoretical framework for
such forces, it becomes possible to make more specific statements
about the nature of their expected couplings, which allows more
focussed analyses of the constraints which can be expected from
these systems.

\subsection{Dark Matter}

If the SLED picture does describe the nature of Dark Energy, then
we must rethink what has become the standard particle-physics
paradigm for the origin of Dark Matter. The standard
(Weakly-Interacting Massive Particle --- WIMP) paradigm involves
the Dark Matter consisting of the relic abundance of a stable
particle having a weak-scale mass and a weak-interaction
annihilation cross section. This is attractive because any such
particle would naturally have the presently-observed abundance
provided only that it were originally in thermal equilibrium with
the other observed particles in the very early universe.

The standard paradigm can be even more ambitious if one assumes
that physics at the TeV scale is governed by the MSSM or one of
its less-minimal variants, since in this case there is a very
natural candidate for such a stable and weakly-interacting
particle having a weak-scale mass. This is because these are
properties which are naturally true for the lightest
supersymmetric particle (LSP). The LSP is expected to be stable on
quite general grounds if the supersymmetric theory enjoys an exact
or approximate $R$ parity, and such a parity seems to be required
in these models in order to explain the non-observation of baryon-
and lepton-number violating decays.

\nisubsubsection{SLED Dark Matter}

\noindent All of this seems to be lost in the MSLED scenario,
since the brane sector is described purely by the Standard Model
and has no super-partners in the usual sense. It is therefore
necessary to think again of where the Dark Matter might arise in
this picture. Although we do not yet know the answer, we wish to
argue that it is again the WIMP proposal which is the most
attractive. There are two reasons for thinking so. The first of
these is the original WIMP motivation: a stable particle with a
weak-scale mass and weak-interaction annihilation cross section
can naturally have the right present-day abundance if it were
originally in thermal equilibrium with ordinary matter at very
high temperatures during the very early universe. (As we discuss
below, this motivation can have additional complications within an
LED framework because of problems which arise if the KK modes in
the two large dimensions are themselves in equilibrium at these
high scales.)

The second reason which points to WIMPs as being the simplest
candidate is given by the `Why Now?' problem of any quintessence
cosmology. This question asks why it should be that the universal
energy density in matter, radiation and Dark Energy should happen
to have been so similar to one another during the very recent
universe. This question is all the sharper given that these three
forms of energy vary very differently as a function of the
universal scale factor as the universe expands.

Although this problem is normally posed for models of Dark Energy
given that the Dark Matter is assumed to be given by some kind of
WIMP, we may ask it in reverse given the SLED picture of what the
Dark Energy is. A clue as to how to do so fruitfully can be found
in ref.~\cite{AHH}, who argue that the Why Now? problem has a
natural solution if two things are true. 1. The Dark Matter is a
relic WIMP abundance, and 2. the Dark Energy is a cosmological
constant which is parameterically of order $\lambda \sim
(M_w^2/M_p)^4$. In this case this reference argues that their
present day abundances would naturally be close to one another
now, since they are both set by ratios of $M_w$ and $M_p$. For
these authors this is true by assumption for the Dark Energy, but
it is also true for WIMPs due to the competition between the
weak-interactions (through their control of the WIMP annihilation
rate) and gravity (through its control of the expansion rate of
the universe).

This picture makes WIMPs attractive within any kind of SLED
proposal, because the Dark Energy density is necessarily predicted
to be of order $\lambda \sim 1/r^4 \sim (M_w^2/M_p)^4$, and so
depends parameterically on $M_w$ and $M_p$ in precisely the
desired way. The question in SLED and MSLED models then becomes
how to identify a promising stable candidate WIMP. This is as yet
unsolved in these models, although it is promising that there can
an enormous number of weakly-interacting states available having
masses at the weak scale. We have seen that for compactifications
from 10 dimensions, string modes, winding modes and Kaluza-Klein
modes for the `other' 4 dimensions can all have masses in the TeV
region.

Two things are required in order to make one of these candidates
into a Dark Matter WIMP. First one needs a good reason why the
state of interest should be stable. This could be done, for
example, for the KK modes of the `other' 4 dimensions if these
dimensions should have isometries giving rise to conserved
charges. For instance, compactification on a 4 torus would provide
the simplest such example, with the Dark Matter being the lightest
KK mode carrying momentum in these weak-scale extra dimensions.
Refs.~\cite{KKDM} consider in detail the relic abundance for
related proposals wherein the lightest KK mode is the Dark Matter
candidate, and find that an acceptable abundance would arise if
the Kaluza-Klein scale were $M_{KK} \sim 2 \pi/\ell \approx 1$
TeV. This fits well into a picture for which $1/\ell \sim 0.1$
TeV, and $M_s \sim (M_g/\ell)^{1/2} \sim 1$ TeV. In view of the
obvious implications for colliders of these scales, it would be
well worth constructing the simplest possible compactification
along these lines in order to investigate its implications in more
detail.\footnote{For toroidal compactifications within string
theory the possibility also arises of using the lightest winding
mode as the Dark Matter particle, and this scenario is $T$-dual to
the situation where the Dark Matter is made up of the lightest KK
mode.}

\nisubsubsection{Over-Closing the Universe}

\noindent It happens that in LED and SLED models the real problem
with Dark Matter is not why there is so much of it, but why there
is so little. This is because LED and SLED models can become
dangerous if KK modes in the {\it large} 2 extra dimensions (as
opposed to the smaller 4 weak-scale dimensions) should be stable
and be in equilibrium at the weak scale. If so, and if some of the
massive KK modes are stable, then these modes become too abundant
and their energy density comes to dominate the universe much too
early in its history.

This problem need not be generic because it depends on more
model-dependent details of the geometry of the extra dimensions
and of the universe's history. When considering LED models it is
normally assumed that these modes were not in equilibrium when
ordinary matter had temperatures above a `normalcy' temperature,
$T_*$, which must be higher than the scale of nucleosynthesis but
cannot be as high as the weak scale. It is more difficult to avail
ourselves of this way out if we wish to explain the Dark Matter
abundance in terms of a thermally generated WIMP, as in the
previous section.

It can also happen that Kaluza-Klein modes do not carry a
conserved charge if the 2 large dimensions are not very symmetric,
since it is the conserved charges associated with
extra-dimensional isometries which the KK modes carry. In this
case KK modes could quickly decay into massless states which would
not dominate the universal energy density. Indeed, it should be
emphasized that it is sufficient to have these extra dimensions
not be symmetric at very early epochs, when the universe's
temperature is above $T_*$, which could be true even if they are
very symmetric right now.

We regard this kind of evolution of the extra dimensions to be
quite likely to occur within the SLED proposal because its energy
cost is not high when temperatures are as large as $T_*$. This is
particularly true, given the propensity of the extra dimensions to
warp in response to changes in tension on the various branes.
(Indeed the possibility that this warping is not small now is the
biggest hurdle which the SLED models must clear in order to
definitively explain the present-day observed Dark Energy
density.) If so, a realistic calculation of the residual energy
tied up in KK modes for the `large' 2 dimensions must await a more
detailed calculation of the cosmological evolution of the 2D
geometry. Such a calculation is not beyond our powers, and is in
progress. It is one more motivation for better understanding the
energetics and dynamics of brane world configurations within the
SLED and LED scenario.

\subsection{Inflation}

Since SLED models require the 6D gravity scale, $M_g$, to be in
the 10 TeV region, it is interesting to speculate whether and how
inflation can arise. We see two main possibilities, depending on
whether or not the fundamental scale $M_g$ itself changes during
the hypothetical inflationary epoch.

\medskip\noindent {\it Weak Scale Inflation}:
It is possible that $M_g$ remains in the TeV range throughout
inflation and its aftermath. If so, then the successful
description of CMB temperature fluctuations would require an {\it
extremely} flat inflaton potential. Since $\delta T/T$ is related
to the inflaton potential $V$ and slow-roll parameter $\epsilon$
by $\delta T/T \propto (V/\epsilon)^{1/2}$, and the observed
fluctuation amplitude requires $(V/\epsilon)^{1/4} \sim 10^{16}$
GeV \cite{LL}, it is clear that if $V^{1/4} \sim M_g \sim 10$ TeV
then we require $\epsilon \sim 10^{-48}$.

\medskip\noindent {\it Dimensional Evolution}:
A more attractive possibility arises if the sizes of the extra
dimensions and the effective 6D gravity scale, $M_g$, themselves
also change during the inflationary epoch, along the lines of what
occurs in the scenarios of ref.~\cite{OrbifoldInflation}. In these
models the internal dimensions evolve like a power of time, $t$,
while the observed 4 dimensions inflate exponentially with $t$.
Such an evolution might ultimately provide an explanation for why
there should be such a large hierarchy amongst the sizes of the
various extra dimensions, as well as potentially providing a
theory of the initial conditions for the moduli of the large 2
internal dimensions.

\section{Have We Been MSLED?}

A great deal of effort has been devoted to exploring the
observable implications of supersymmetry over the last few years.
The combination of good theoretical motivation and distinctive
experimental signatures has attracted the interest of theorists,
phenomenologists, cosmologists and experimentalists alike. The
supersymmetric extension of the Standard Model has emerged from
this study as the conservative choice for the most likely
replacement of the Standard Model itself.

Implicit in this choice is the belief that the supersymmetric
Standard Model is an inevitable consequence of TeV-scale
supersymmetry breaking within a microscopic theory which is
supersymmetric. Although originally motivated by solving the
cosmological constant problem, what is intriguing about the SLED
proposal is that it provides a radically different way in which
supersymmetry could be realized at low energies. As such it
provides a counterexample to the statement that the supersymmetric
Standard Model is the sole low-energy manifestation of TeV-scale
supersymmetry breaking.

Equally important, the connection with the cosmological constant
makes the SLED proposal unique in several ways. First, if proven
to be successful, it would be the only extant proposal which can
provide a natural explanation for the observed size of the dark
energy. Second, since it involves modifying physics at extremely
low energies, it has very many observational consequences for
particle physics and astrophysics, apart from its predictions for
the dark energy. Finally, by relating the size of the extra
dimensions to the observed dark energy, it is very predictive
because it removes the freedom to shrink these dimensions to evade
their experimental consequences.

In this article we provide a first step towards exploring the
phenomenology of SLED models. We do so by identifying several
robust consequences of the SLED proposal, as well as proposing a
naturally minimal special case, MSLED, for which even more
specific predictions may be made. The model which results
encompasses many of the main properties of the large extra
dimensions scenarios, but with a more focussed set of predictions
because of the inability to tweak the model by choosing more large
dimensions, or to arrange the dimensions to be smaller or to
significantly change the field content and interactions in the
6-dimensional bulk.

Among the observable consequences which follow quite robustly for
the SLED proposal are:
\begin{itemize}
\item Deviations from Newton's inverse-square law {\it must} occur
at $\mu$m scales.
\item The Dark Energy should be capable of dynamically evolving
even during the present epoch, due to the presence of at least one
very light scalar corresponding to changes in the size of the
extra large dimensions.
\item The existence of the cosmologically light scalar(s) predicts
that gravity is described by a scalar-tensor model over
astrophysical distance scales.
\item Astrophysical systems (like supernovae) and experiments at
TeV scales should see (or be close to seeing) significant missing
energy processes corresponding to emission of Kaluza Klein modes
in the large two dimensions. Because the extra dimensions are
supersymmetric, in SLED there are significantly more such states
than there are gravitons.
\end{itemize}

The minimal MSLED version of the model also predicts:
\begin{itemize}
\item Particles on our brane consist purely of those of the
Standard Model itself. In particular the Standard Model particles
have {\it no superpartners} in the effective theory below the TeV
scale where the scale of 6D gravity becomes important. They do not
because supersymmetry is badly broken (at the TeV scale) on our
brane.
\item Low-energy particle physics should be well-described by the
Standard Model, with suppressed flavour-changing interactions.
Since the influence of extra-dimensional physics is suppressed at
lower energies, it is unlikely that new effects should emerge in
experiments involving low-energy $K$ and $B$ mesons.
\end{itemize}

Many of these implications are at the edge of being tested
experimentally, both from accelerator and table-top experiments,
so if the SLED picture is right we are likely to know within the
comparatively near future.

\subsection{MSLED {\it vs} the MSSM}

Since MSLED is being proposed as a low-energy realization of
supersymmetry which is a well-motivated and concrete alternative
to the standard MSSM picture, it is perhaps worth comparing their
relative merits and liabilities given our present state of
knowledge.

\medskip\noindent {\it Theoretical Motivation:}
One of the main practical benefits of the supersymmetric Standard
Model has been that it is both theoretically well-motivated, and
yet is also concrete and broadly predictive, particularly in its
MSSM form. This makes it possible to test in detail, and compare
with the predictions of the Standard Model
itself.\footnote{Indeed, if naturality issues are taken seriously
the MSSM is so predictive as to be close to being ruled out.}
MSLED is also very predictive, and has two strong theoretical
motivations. Its main motivation comes from the SLED description
of the cosmological constant, but it also shares the theoretical
motivation of describing a class of weak-scale supersymmetry
breaking which can plausibly arise from a more fundamental
microscopic theory like string theory.

\medskip\noindent {\it Hierarchy Problem:}
The MSSM and MSLED both address the hierarchy problem, although in
different ways. The MSSM provides a way to understand the natural
stability of the hierarchy against loop corrections, but does not
in itself provide an explanation of why the force of gravity
should be weak compared with the electro-weak force in the first
place. (The explanation of this requires a theory of how
supersymmetry breaks in a more microscopic context, which is not
yet available.) In MSLED the TeV scale is fundamental and the weak
strength of gravity relies on the size of the 2 large extra
dimensions being so large, and the explanation of why this is true
becomes a dynamical question. Given that this size is large,
MSLED's key claim is that it can preserve the resulting hierarchy
against the influence of loops. The main issue to be addressed in
showing this to be true also involves dynamics: do the extra
dimensions warp as higher-energy scales are integrated out?

In both cases the explanation for the difference between the
electro-weak and Planck scales could be dynamical. In the MSSM it
depends on the way supersymmetry is broken, possibly by
non-perturbative effects that lead naturally to a large hierarchy
of scales. In the MSLED case it depends on the mechanism that
fixes the size and shape of the extra dimensions, such as in the
scalar potential discussed in section 2. An important point for
the MSLED is that the same scale that is needed to understand the
electro-weak hierarchy problem, also provides the natural scale
for dark energy.

\medskip\noindent {\it Low-Energy Supersymmetry:}
Both MSSM and MSLED involve supersymmetry broken at the
electro-weak scale, although the way supersymmetry is realized in
the low-energy theory is very different. The field content of the
MSSM fills out complete 4D supermultiplets, and so supersymmetry
is realized linearly. This happens because the typical splittings
of masses in supermultiplets is smaller than the cutoff at the TeV
scale. By contrast, in MSLED since supersymmetry is broken on the
brane at the TeV scale, there are no super-partners for ordinary
particles in the low-energy theory. Supersymmetry is therefore
only realized nonlinearly there \cite{nonlinearbranesugra}. The
bulk physics, however, has a small supersymmetry-breaking scale,
and so particles fall into 6D supermultiplets which linearly
realize 6D supersymmetry.

\medskip\noindent {\it Dark Energy:}
The MSSM does not explain the small observed size of the
cosmological constant within the low-energy effective theory, and
so does not protect its value to be smaller than of order $(1 {\rm
TeV})^4$. One of the main motivations for the MSLED proposal is
its potential to explain this major problem in a natural way.
Being a much more recent proposal, it is not completely clear as
yet whether the MSLED proposal will provide the definitive
solution (with the main obstacle likely to be the possibility of
the large 2 dimensions warping too much in response to the various
internal brane tensions). Nevertheless we believe it goes further
in this regard than does the MSSM, inasmuch as it provides a
mechanism for understanding why ordinary particles like the
electron do not in themselves provide too large a cosmological
constant.

\medskip\noindent {\it Gauge Coupling Unification:}
The measured strengths of the electro-weak and strong interactions
at the weak scale appear to be consistent with the unification of
these three gauge coupling constants, once they are run up to much
higher energy scales --- subject to the assumption of a plausible
spectrum of super-partners at the TeV scale. The ability to
understand this running is often considered as a triumph of the
MSSM. The MSLED scenario does not offer a similarly natural
explanation for this apparent coupling unification at the GUT
scale, since it is the TeV scale rather than a larger scale which
is fundamental within this model. One attitude to take within
MSLED is that the evidence for unification is not that compelling,
and so to simply put it aside.

An alternative approach is also possible, however
\cite{HDUnification,costas, luis}. For instance in \cite{costas},
it is argued that precisely in the case of two large extra
dimensions, there is a natural logarithmic `running' of the gauge
couplings coming from the logarithmic dependence on position of
the massless propagator in two dimensions. This assumes that gauge
couplings are position dependent because they are given by
expectation values of bulk fields (such as happens in string
theory) that vary logarithmically with position in the extra
dimensions. In this picture, some bulk fields (like twisted
moduli) couple differently to different gauge groups, and the
place in the bulk where they vanish would correspond to the
unification point. (The large energy desert over which couplings
run is replaced in this picture by the large distance between our
brane and this unification point within the bulk.) Although it is
far from clear that such a picture can be consistent with the SLED
proposal, it is a logical possibility which can be explored.

\medskip\noindent {\it Proton Stability:}
The Standard Model beautifully explains the baryon and lepton
number conservation of the renormalizable interactions as being an
accidental consequence of the model's particle content and gauge
symmetries. Proton stability is then naturally assured if the
scale which suppresses any higher-dimension non-renormalizable
effective interactions is of order the GUT scale, since the first
baryon-number violating interactions arise at dimension 6. This
property is lost in the MSSM, for which gauge invariance permits
many baryon and lepton-number violating renormalizable
interactions. These must be forbidden using extra symmetries like
$R$-parity, and even then care must be taken to avoid having
dangerous dimension-5 interactions contribute to proton decay.
MSLED inherits the natural explanation for lepton- and
baryon-number conservation for all renormalizable interactions
since it is simply given on our brane by the Standard Model
itself. Just as for the Standard Model, baryon-number violating
interactions first arise at dimension 6, but for MSLED the natural
scale which suppresses these interactions is only $M_g \sim 10$
TeV. Consequently, the existence of a symmetry (like baryon-number
itself) must also be invoked to adequately suppress those
non-renormalizable interactions which could mediate proton decay.
Indeed, gauged baryon-number symmetries frequently arise within
explicit quasi-realistic brane-world string compactifications
\cite{BinBW, real2}.

\medskip\noindent {\it Flavour Problems:}
The MSSM has many soft supersymmetry-breaking parameters, which
potentially include many arbitrary phases and flavour-changing
interactions. The current lack of experimental evidence for these
interactions introduces a new naturalness problem whose resolution
is not provided within the model. New problems of this sort do not
arise for MSLED to the extent that it is only the Standard Model
which appears on our brane. (Of course, MSLED does not in itself
shed any light on why the known fermions have the particular
masses and mixing angles they are observed to have.)

\medskip\noindent {\it Falsifiability:}
Both the MSSM and MSLED are likely to be decisively tested (and
possibly ruled out) in the next few years. Their experimental
signatures are quite different, with the MSSM predicting a rich
spectroscopy of supersymmetric particles, and MSLED predicting
very finely-spaced Kaluza-Klein states and missing energy
 at accelerators, as well as deviations of
gravity in table-top experiments. Depending on how it is embedded
into a more fundamental theory it may also predict that TeV-scale
KK modes and string states are also present to be seen in
colliders.

\medskip\noindent {\it Plan B:}
Both models have simple extensions that share their main
attractive features and predictions but which can provide some
(but not all) experimental signatures, should these be required
(such as by experiments searching for flavour changes in meson
physics). In the MSSM new gauge and matter supersymmetric
multiplets could be considered as well as more general soft
supersymmetry breaking couplings. In the MSLED case, various
extensions of the standard model could be considered on our brane
rather than just the Standard Model. These extensions could
eventually be explored if required either on experimental or
theoretical grounds.

\medskip\noindent {\it Dark Matter:}
Both models favour a WIMP candidate for dark matter, although the
MSSM has in the lightest super-partner a very generic candidate,
due to the assumed conservation of $R$-parity. MSLED may also have
many possible candidates, such as higher-dimensional KK modes,
although further studies are required in order to better identify
the alternatives.

\medskip\noindent {\it Fundamentalism:}
Neither the MSSM nor MSLED have been derived from a more
fundamental theory such as string theory, but both have reasonable
prospects for being found near already known string vacua.

\medskip
There are clearly more open questions than answers regarding the
possible experimental tests of the SLED and MSLED scenarios, but
we regard the exploration of these issues to be worth pursuing
given the exciting connections which may emerge between particle
physics, gravitation, astrophysics and cosmology.

\section*{Acknowledgements}

We thank useful conversations with C. Csaki, E. Dudas, D.
Ghilencea, E. Kiritsis, G. Moore, A. Nelson, M. Petrini, S.
Parameswaran, A. Pomarol, L. Randall, A. Tomasiello and A.
Zaffaroni. FQ, CB and JM respectively thank the Ecole
Polytechnique, IFAE, Barcelona and McGill University,  for
hospitality while this work was being completed. C.B.'s research
has been supported by grants from NSERC (Canada), FQRNT
(Qu{\'e}bec) and McGill University. FQ is partially supported by
PPARC and a Royal Society Wolfson award. J.M receives financial
support from the Ram\'on y Cajal Program and grant FPA2002-00748.

\end{document}